\documentclass[twocolumn,showpacs,preprintnumbers,amsmath,amssymb,prb]{revtex4}
\usepackage{graphicx}
\usepackage{dcolumn}
\usepackage{bm}
\usepackage{natbib}
\begin{document}
\title{Magnetic properties of Gd$_x$Y$_{1-x}$Fe$_2$Zn$_{20}$: dilute, large, $\textbf {S}$ moments in a nearly ferromagnetic Fermi liquid}
\author{S. Jia, Ni Ni, S. L. Bud'ko, P. C. Canfield}
\affiliation{Department of Physics and Astronomy, Iowa State University, Ames, Iowa 50011, USA\\
Ames laboratory, USDOE, Ames, Iowa 50011, USA\\}

\begin{abstract}
Single crystals of the dilute, rare earth bearing, pseudo-ternary series, Gd$_x$Y$_{1-x}$Fe$_2$Zn$_{20}$ were grown out of Zn-rich solution. 
Measurements of magnetization, resistivity and heat capacity on Gd$_x$Y$_{1-x}$Fe$_2$Zn$_{20}$ samples reveal ferromagnetic order of Gd$^{3+}$ local moments across virtually the whole series ($x \geq 0.02$). 
The magnetic properties of this series, including the ferromagnetic ordering, the reduced saturated moments at base temperature, the deviation of the susceptibilities from Curie-Weiss law and the anomalies in the resistivity, are understood within the frame work of dilute, $\textbf {S}$ moments (Gd$^{3+}$) embedded in a nearly ferromagnetic Fermi liquid (YFe$_2$Zn$_{20}$).
The s-d model is employed to further explain the variation of $T_{\mathrm{C}}$ with $x$ as well as the temperature dependences of of the susceptibilities.

\end{abstract}


\maketitle
\section{Introduction}

Materials that are just under the Stoner limit manifest large electronic specific heat and enhanced paramagnetism and are sometimes known as nearly ferromagnetic Fermi liquids (NFFL)\cite{moriya_spin_1985,brommer_strongly_1990}.
Archetypical examples, such as Pd\cite{jamieson_magnetic_1972}, Ni$_3$Ga\cite{de_boer_high_1967}, TiBe$_2$\cite{matthias_itinerant_1978} and YCo$_2$\cite{lemaire_magnetic_1966}, have been studied for several decades.
In addition to the interesting, intrinsic properties of these compounds, the introduction of local moments into these highly polarizable hosts has lead to both experimental\cite{nieuwenhuys_magnetic_1975} and theoretic interest\cite{larkin_magnetic_1972,maebashi_singular_2002}.
In such highly polarizable hosts, local moment impurities can manifest long range, ferromagnetic order even for very low concentrations (0.5 at.\% Fe in Pd\cite{mydosh_magnetic_1968} and 1 at.\% Gd in Pd\cite{crangle_ferromagnetism_1964}). 

Recently, YFe$_2$Zn$_{20}$ was found to be a ternary example of a NFFL with a Stoner parameter $Z \sim  0.9$,\cite{jia_nearly_2007} as compared to $Z \sim  0.83$ for Pd, indicating strongly correlated electron behavior.
When the large, $\textbf {S}$ moment bearing, Gd$^{3+}$ replaces the non-magnetic Y$^{3+}$ ions, it was found that GdFe$_2$Zn$_{20}$ has a remarkably high ferromagnetic Curie temperature($T_{\mathrm{C}}$) of 86 K. 
Both of these compounds belong to the much larger, isostructural RT$_2$Zn$_{20}$ (R = rare earth, T = transition metal such as Fe, Co, Ni, Ru, Rh, Os, Ir, Pt)\cite{nasch_ternary_1997,torikachvili_six_2007} family, in which the R and T ions each occupy their own unique, single, crystallographic sites.
In these dilute, rare earth bearing, intermetallic compounds (less than 5 at.\% rare earth), the R ions are fully surrounded by Zn nearest and next nearest neighbors to form a Frank-Kasper-like Zn polyhedron; the T site is also surrounded by a nearest and next nearest neighbor, Zn shell. 
The shortest R-R spacing is $\sim$6 {\AA}. 
Motivated by these intriguing magnetic and structural properties, we focus, in this work, on the pseudo-ternary series Gd$_x$Y$_{1-x}$Fe$_2$Zn$_{20}$, which can be used as a model for studying the effects of titrating very dilute local moments into a nearly ferromagnetic Fermi liquid. 
Given that RFe$_2$Zn$_{20}$ is a dilute, rare earth bearing intermetallic, dilution of Gd onto the Y site (i) changes the lattice parameter by less than 0.2 \%, (ii) does not change the band filling, (iii) does not change the all Zn local environment of either the Gd or Fe ions, and (iv) allows for the dilution of Gd in the system to be studied down to $x\approx 0.005$, i.e. down to approximately 200 $p.p.m.$ Gd. 
As shown below, single crystals of Gd$_x$Y$_{1-x}$Fe$_2$Zn$_{20}$ can be easily grown by a Zn, self flux method\cite{jia_nearly_2007,canfield_growth_1992}, and the Gd concentration can be consistently inferred via a variety of methods.

In this paper, we report on the characterization of single crystals of Gd$_x$Y$_{1-x}$Fe$_2$Zn$_{20}$ by X-ray diffraction, Energy Dispersive X-ray Spectroscopy (EDS), magnetization, resistivity and heat capacity measurements. These data reveal ferromagnetic order of the Gd$^{3+}$ local moment above 1.80 K for Gd concentration above $x=0.02$. These results will be discussed within the framework of the so-called s-d model\cite{shimizu_itinerant_1981}, based on the mean field approximation, and used to explain the variation of $T_{\mathrm{C}}$ across the series with respect to x.

\section{Experimental Methods}
\begin{figure}
  \begin{center}
  \includegraphics[clip, width=0.45\textwidth]{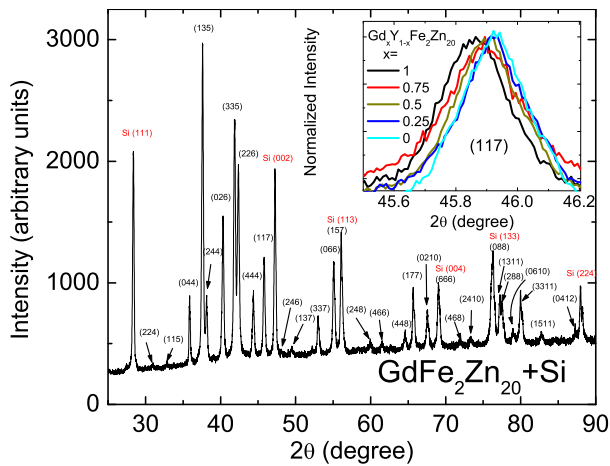}\\
  \caption{Powder X-ray diffraction pattern of GdFe$_2$Zn$_{20}$ with a Si internal standard (using Cu K$_\alpha $ radiation) with main peaks indexed. Inset: the normalized intensity of the (117) peak of Gd$_x$Y$_{1-x}$Fe$_2$Zn$_{20}$ for representative $x$ values, with the positions calibrated by the nearby Si(002) peak.}
  \label{fig:X-ray}
  \end{center}
\end{figure}

Single crystals of Gd$_x$Y$_{1-x}$Fe$_2$Zn$_{20}$ were grown from a Zn-rich self flux\cite{jia_nearly_2007,canfield_growth_1992}.
For $x>0.02$, high purity elements were combined in a molar ratio of (Gd$_x$Y$_{1-x}$)$_2$Fe$_4$Zn$_{94}$. 
For $x$ less than 0.02, a Y$_{0.9}$Gd$_{0.1}$ master alloy was made via arc melting and appropriate amounts of this alloy were added to elemental Y to reduce the uncertainties associated with weighing errors.
The constituent elements (or alloy) were placed in a 2 ml, alumina crucible and sealed in a silica tube under approximately 1/3 atmosphere of high purity Ar (used to help reduce the evaporation and migration of zinc during the growth process) and then heated up to 1000 $^\circ$C and cooled, over a period of 80 h, to 600 $^\circ$C, at which point the remaining liquid was decanted.
Growths such as these often had only a few nucleation sites per crucible and yielded crystals with typical dimensions of at least $7\times 7\times 7$  $mm^3$.
Residual flux and/or oxide slag on the crystal surfaces was removed by using 0.5 vol.\% HCl in H$_2$O in an ultrasonic bath for 1--2 h.
The samples were characterized by room temperature powder X-ray diffraction measurements using Cu K$_\alpha $ radiation with Si ($a=5.43088 ~\AA$) as an internal standard in a Rigaku Miniflex powder diffractometer(Fig. \ref{fig:X-ray}).
The Rietica, Rietveld refinement program was employed to obtain the lattice constants, which vary linearly for $1\leq x \leq 0$.
This shift can be seen in the $(117)$ peak position for selected $x$ values (see Fig. \ref{fig:X-ray}, inset).
EDS measurements were made in a JEOL model 5910lv-SEM with a Vantage EDS systerm for representative samples. 

In order to measure the electrical resistivity with a standard AC, four-probe technique, the samples were cut into bars using a wire saw.
The bars typically had lengths of 2--4 mm parallel to the crystallographic [110] direction, and widths and thicknesses between 0.2--0.4 mm.
Electrical contact was made to these bars by using Epo-tek H20E silver epoxy, with typical contact resistances of about 1 Ohm.
AC electrical resistivity measurements were performed with $f=16$ Hz and $I= 1-0.3$ mA in a Quantum Design PPMS-14 or PPMS-9 instrument ($T=1.85 - 310$~K). 
Temperature dependent specific heat measurements were also performed by using the heat capacity option of these Quantum Design instruments, sometimes using the $^3$He option.

\begin{figure}
  \begin{center}
  \includegraphics[clip, width=0.45\textwidth]{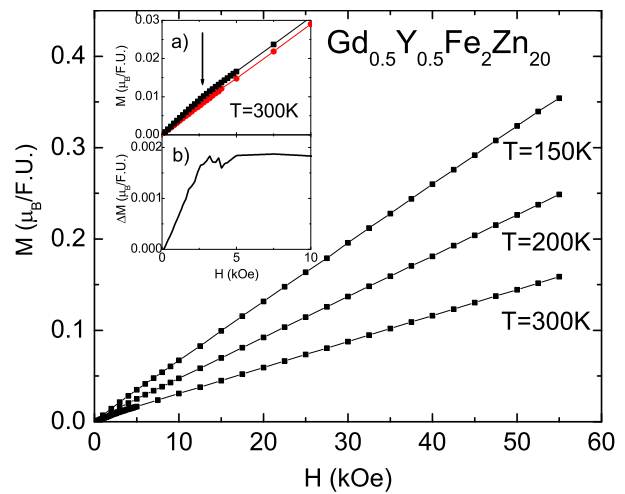}\\
	\caption{Magnetization $M$ with respect to applied field $H$ for a sample of Gd$_{0.5}$Y$_{0.5}$Fe$_2$Zn$_{20}$ at 150~K, 200~K and 300~K. The solid lines are guides to the eye. Inset a: detailed magnetization of two samples of Gd$_{0.5}$Y$_{0.5}$Fe$_2$Zn$_{20}$ at 300~K. The data shown in black (same data as in main figure) has slope change feature (indicated by an arrow); while the data shown in red does not. Inset b: the difference of the black and red reveals the saturation of ferromagnetic impurity above 5~kOe.}
  \label{fig:MH50}
	\end{center}
\end{figure}

DC magnetization was measured in a Quantum Design superconducting quantum interference device (SQUID) magnetometer, in a variety of applied fields ($H \leq 55$ kOe) and temperatures (1.85 K $\leq  T \leq$ 375 K).
In some crystals, the magnetization with respect to magnetic field measurements at 300 K showed a slight non-linearity with a small slope change around 3 kOe (Fig. \ref{fig:MH50}).
This specific behavior is believed to be due to a small amount of ferromagnetic impurity, possibly Fe or FeO$_x$ ($2\times 10^{-5}$ $\mu _B$/mol to $2\times 10^{-3}$ $\mu _B$/mol) on the crystal.
This feature is most likely extrinsic because the extent of the slope change is sample-dependent; some samples showing no feature at all (inset of Fig. \ref{fig:MH50}).
This feature is most clearly seen when two samples from the same batch (one with feature, one without) are compared (Fig. \ref{fig:MH50}, inset a) or even subtracted from each other (Fig. \ref{fig:MH50}, inset b).
Given that this small, extrinsic ferromagnetic contribution saturates by $H\approx 10$ kOe (Fig. \ref{fig:MH50}, inset b), the high temperature susceptibility can be determined by $\chi (T)=\frac{\Delta  M}{\Delta H}=\frac{M_{(H=50kOe)}-M_{(H=20kOe)}}{30kOe}$. 
In this temperature region the intrinsic magnetization is a linear function of applied magnetic field for $20$ $kOe\leq H\leq 50$ $kOe$ (Fig. \ref{fig:MH50}).
At lower temperatures, closer to $T_{\mathrm{C}}$, the sample's intrinsic magnetization become large enough that we can measure $\chi (T)$ directly as $M/H$ for $H=1$ kOe.

\section{Experiments Results}

\begin{figure}
  \begin{center}
  \includegraphics[clip, width=0.45\textwidth]{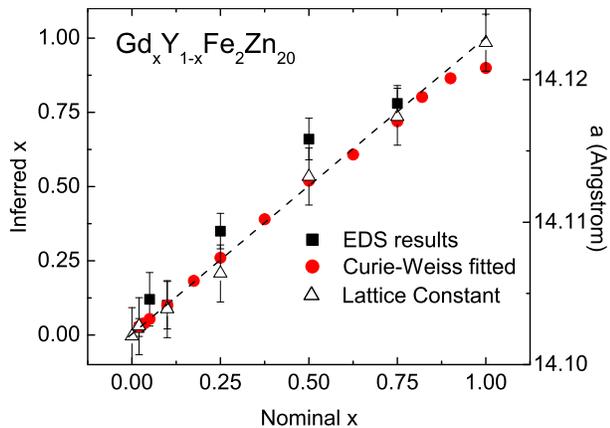}\\
  \caption{Gd concentration inferred from EDS (solid squares) and high temperature magnetic susceptibility (solid circles). The open triangles represent lattice constants. The dash line is location where inferred $x$ equals nominal $x$ and also represents a linear dependance of the lattice parameter.}
  \label{fig:EDS}
  \end{center}
\end{figure}

The size of the cubic unit cell, as determined by powder X-ray diffraction measurements, shows a linear dependence on $x$ as it is varied from 0 to 1 (Fig. \ref{fig:EDS}).
The error bars of the lattice constants were estimated from the standard deviation determined by measurements on three samples from the same batch. 
These data are compliant with Vegard's law and imply that the nominal $x$ is probably close to the actual $x$.

In order to check this further, EDS was used. This is a direct method to determine $x$, although it loses some of its accuracy because of the low, total rare earth concentration ($< 5$ at.\%). 
Nevertheless, several representative members of the Gd$_x$Y$_{1-x}$Fe$_2$Zn$_{20}$ series were measured and the inferred $x$ values are close to the nominal $x$ values within the fairly large error bars (Fig. \ref{fig:EDS}).  

\begin{figure}
  \begin{center}
  \includegraphics[clip, width=0.45\textwidth]{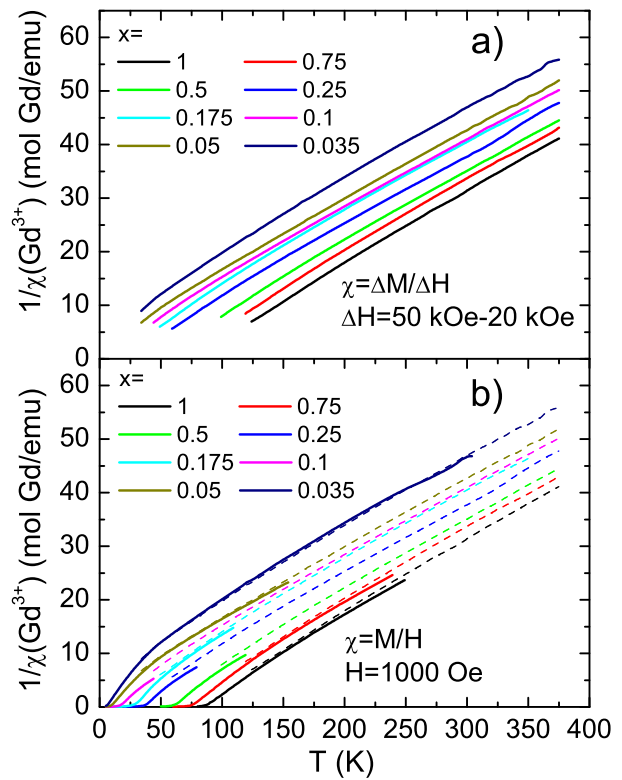}\\
  \caption{ $1/\chi _{Gd^{3+}}$ vesus temperature for representative members of the Gd$_x$Y$_{1-x}$Fe$_2$Zn$_{20}$ sereies. Note: data is normalized to mole Gd using $x$ inferred from high-temperature data. (a): obtained under high magnetic field. (b) Solid lines: obtained under 1~kOe applied field; dash lines: under high magnetic field.}
  \label{fig:HM}
  \end{center}
\end{figure}

Another way to estimate the concentration of gadolinium in the grown crystals is based on the analysis of the high temperature magnetic susceptibility data, which can be expressed as:
\begin{equation}
\chi _{Gd_{x}Y_{1-x}Fe_{2}Zn_{20}}=\chi _{Gd^{3+}}+\chi _{YFe_{2}Zn_{20}}\
\label{eqn:1}
\end{equation}
Experimentally, $\chi _{Gd^{3+}}$ obeys the Curie-Weiss law above 150 K (Fig. \ref{fig:HM}a), from which the paramagnetic Curie temperature $\theta_C$ and Curie constants $C$ can be extracted.
The value of $x$ can be inferred by fixing the effective moment of Gd$^{3+}$ as 7.94 $\mu _B$.
These values of inferred $x$ are also plotted in Fig. \ref{fig:EDS}.
The agreement between each of these three different methods of determining inferred $x$ and the nominal $x$ value is good and for the rest of this paper nominal values will be used to estimate actual Gd content.

Another aspect of Fig. \ref{fig:HM} that is noteworthy is that all $\chi _{Gd^{3+}}$ data sets deviate from their high temperature Curie-Weiss behaviors as the system approaches the magnetic ordering temperature.
Since high fields can shift and broaden the features associated with ferromagnetism, at lower temperatures a field of 1 kOe was used (Fig. \ref{fig:HM}b).
Whereas this deviation cannot be associated with the formation of superparamagnetic clusters above $T_{\mathrm{C}}$ (this would cause a slope change toward the horizontal rather than toward the vertical), it can be understood in terms of an increasing coupling between the Gd$^{3+}$ local moments associated with the strongly temperature dependent, polarizable electronic background of the YFe$_2$Zn$_{20}$ matrix\cite{jia_nearly_2007}(see discussion below).

\begin{figure}
  \begin{center}
  \includegraphics[clip, width=0.45\textwidth]{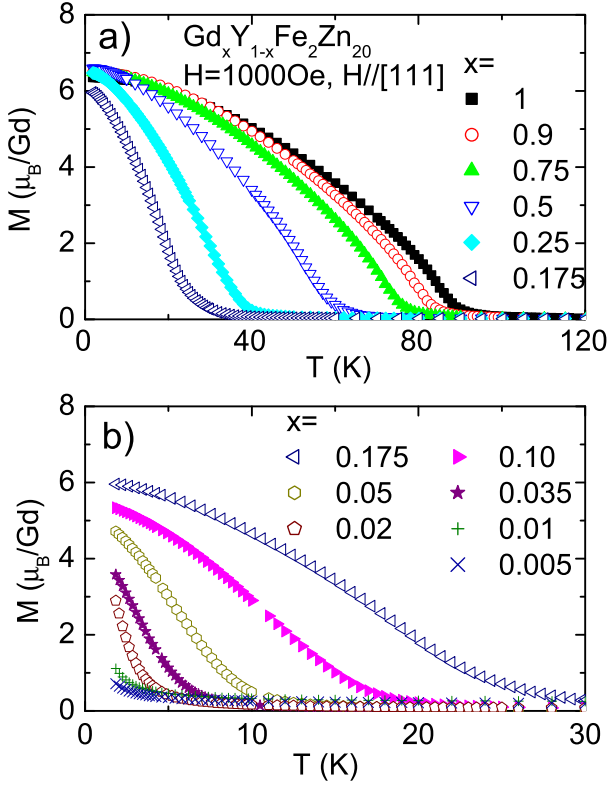}\\
  \caption{Temperature dependent magnetization of Gd$_x$Y$_{1-x}$Fe$_2$Zn$_{20}$, H = 1000 Oe, for (a) $1.0\geq x\geq 0.175$, (b) $x\leq 0.175$.}
  \label{fig:MTall}
  \end{center}
\end{figure}

Figure \ref{fig:MTall} shows the temperature dependent magnetization in an external field $H=1000$ Oe for the whole range of $x$ values.
Ferromagnetic ordering can be clearly seen below 90 K for $x=1$.
This ordering temperature decreases monotonically as $x$ decreases, although the exact values of $T_{\mathrm{C}}$ can not be unambiguously inferred from these plots.
For $x\leq 0.035$, it becomes difficult to determine whether the compounds manifest ferromagnetism above the base temperature (1.85 K) based on the $M(T)$ curves alone.
Even at 1000 Oe, for $x\geq 0.25$, the low-temperature magnetization is just slightly below the Hund's ground state value 7 $\mu _B$/Gd  at the base temperature (Fig. \ref{fig:MTall}a).
For $x<0.25$ the low temperature, $H=1000$ Oe, magnetization decreases with decreasing $x$ (Fig. \ref{fig:MTall}b).

\begin{figure}
  \begin{center}
  \includegraphics[clip, width=0.45\textwidth]{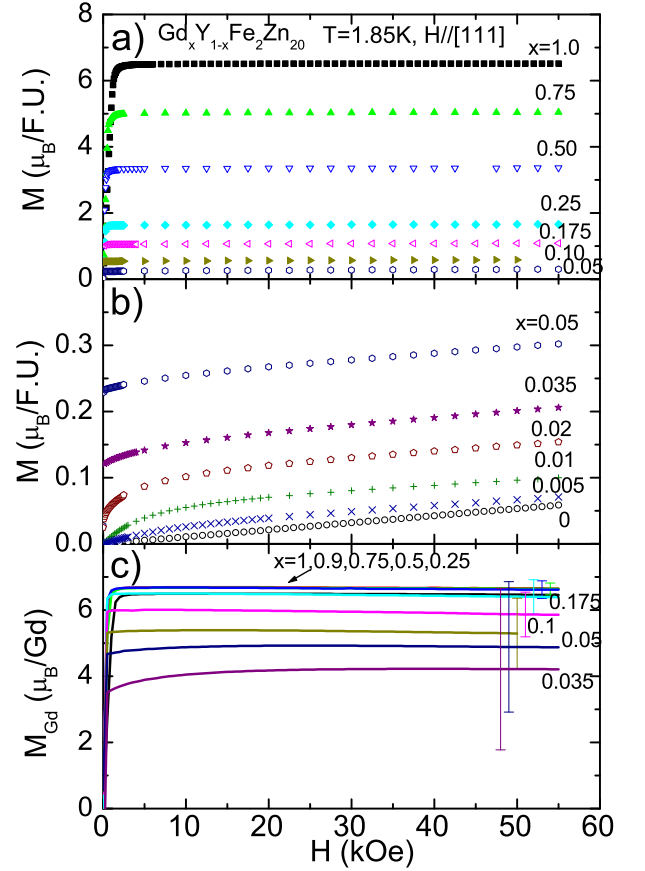}\\
  \caption{(a) and (b) Field dependent magnetization of Gd$_x$Y$_{1-x}$Fe$_2$Zn$_{20}$ at 1.85 K. (c) Field dependent magnetization of Gd$_x$Y$_{1-x}$Fe$_2$Zn$_{20}$ at 1.85 K, normalize to Gd$^{3+}$ content (see text).The error bars were estimated by allowing for a $\pm 0.02$ variation of $x$}
  \label{fig:MHall}
  \end{center}
\end{figure}

Field-dependent magnetization measurements were made for each sample at base temperature (Fig. \ref{fig:MHall}).
For compounds with $x\geq 0.035$, the magnetization rapidly saturates as the magnetic field increases, consistent with a ferromagnetic ground state at 1.85 K.
For $x\leq 0.01$, the $M(H)$ curves vary more smoothly with H and  are more consistent with a paramagnetic state at 1.85 K.
The $x=0.02$ data are more ambiguous and require a still more detailed analysis. 

For $H>10$ kOe the $M(H)$ data for $x\leq 0.05$ vary approximately linearly with $H$ and have slopes comparable to that of YFe$_2$Zn$_{20}$.(Fig. \ref{fig:MHall}b)
For all $x$ values the magnetization can be thought of as a combination of the magnetization of Gd$^{3+}$ ions and the highly polarizable background.
In order to extract the magnetization of the Gd$^{3+}$ ions, a background of $M_{YFe_2Zn_{20}}$ was subtracted from the $M(H)$ data.
The $M_{Gd}(H)$  data are plotted in Fig. \ref{fig:MHall}c normalized to the nominal x values.
For $x\geq 0.25$ the saturated magnetization is essentially constant with a value slightly less than 7 $\mu _B$/Gd.\cite{jia_nearly_2007} 
For $x<0.25$ there is an apparent decrease in the saturated magnetization with decreasing $x$, but it should noted that the error bars, coming from the estimated  $\pm 0.02$ uncertainty of $x$, increase with decreasing $x$.
These increasing error bars make it unclear whether the saturated moment of the Gd impurities is constant or decreasing in the small $x$ limit. 

\begin{figure}
  \begin{center}
  \includegraphics[clip, width=0.45\textwidth]{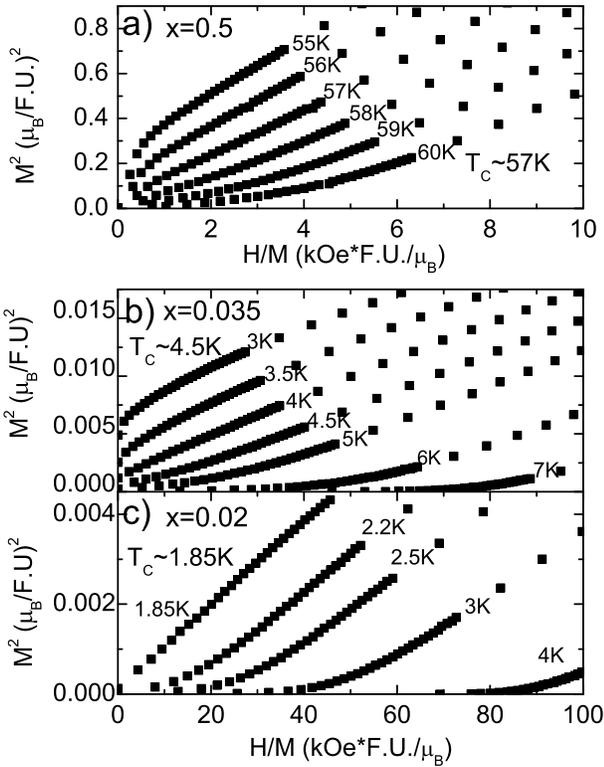}\\
  \caption{Arrott plots for representative members of the Gd$_x$Y$_{1-x}$Fe$_2$Zn$_{20}$ series: $x$ = (a) 0.5, (b) 0.035 and (c) 0.02.}
  \label{fig:Arrott}
  \end{center}
\end{figure}

A fuller analysis of $M(H)$ data, particularly the analysis of magnetization isotherms known as Arrott plots\cite{arrott_criterion_1957}, at a set of temperatures near $T_{\mathrm{C}}$ has been found to be a useful, and in some case even the best method to determine $T_{\mathrm{C}}$ for the $x<0.25$ samples.
The method is based on the mean field theory, in which $M^2$ is linear in $I/M$ with zero intercept at the critical temperature $T_{\mathrm{C}}$, where $I$ is the internal field, equal to the difference between the external, applied field $H$ and the demagnetizing field $D_m$.
For an ellipsoid of Gd$_x$Y$_{1-x}$Fe$_2$Zn$_{20}$, the demagnetizing field equals\cite{chikazumi_physics_1997}:
\begin{equation}
D_{m}=M\cdot D\cdot \frac{4\pi }{(14\AA )^{3}}\cdot N_{A}=0.061\cdot D\cdot M
\label{eqn:2}
\end{equation}
where $M$ is the magnetization (emu/mol); $D$ is a geometric factor that can range from 1 to 0, and $N_A$ is Avogadro number.
Thus $I/M$, in units of $\mu _B$/kOe , is:
\begin{equation}
\frac{I}{M}=\frac{H-D_{m}}{M}=\frac{H}{M}-0.34\times D\
\label{eqn:3}
\end{equation}
Using $H$, instead of $I$ , in Arrott plots will shift the data along $H/M$ axis in the positive direction by $0.34\times D$, which would experimentally introduce an error in the value of $T_{\mathrm{C}}$ for a flat shaped sample ($D\sim 1$) of GdFe$_2$Zn$_{20}$.
Nevertheless, even in this extreme case, this error drops as $x$ decreases due to reduction of the samples' magnetization as Gd$^{3+}$ is diluted out (notice the different scale of the $M$ axis for $x<0.05$ in Fig. \ref{fig:Arrott}).
Due to these concerns, rod-like-shape samples were measured along their long axis for the magnetization isotherms for samples with $x>0.5$.
This shape ensures $D$ is minimized. 
Figures \ref{fig:Arrott}a and b show $T_{\mathrm{C}}=57\pm 0.5$~K for $x=0.5$ and $T_{\mathrm{C}}=4.5\pm 0.5$~K for $x=0.035$ respectively.
For $x=0.02$, Fig. \ref{fig:Arrott}c shows $T_{\mathrm{C}}=1.85$~K, a result that helps explain the difficulty experienced in determining the base-temperature magnetic state based on the $M(T)$ and $M(H)$ data discussed above.
The $T_{\mathrm{C}}$ values determined for the Arrot plot analysis for all $x$ are shown below in Fig. \ref{fig:summary}. 

\begin{figure}
  \begin{center}
  \includegraphics[clip, width=0.45\textwidth]{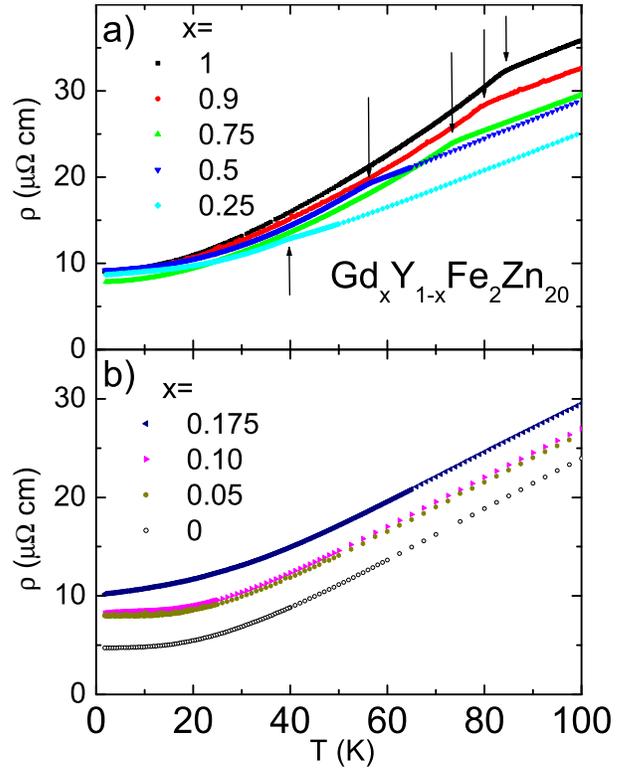}\\
  \caption{Zero-field resistivity for current along the [110] direction. The arrows represent $T_{\mathrm{C}}$ determined from Arrott plot analyses.}
  \label{fig:RT}
  \end{center}
\end{figure}

The temperature dependent electric resistivity data, $\rho (T)$ (measured in zero applied magnetic field), of the Gd$_x$Y$_{1-x}$Fe$_2$Zn$_{20}$ compounds are shown, for representative $x$ values, in Fig. \ref{fig:RT}. 
For $x\geq 0.25$, $\rho (T)$ curves show a kink at $T_{\mathrm{C}}$ due to the loss of spin disorder scattering below this temperature.
In contrast, for $x\leq 0.175$, no clear kink can be detected.
$T_{\mathrm{C}}$ values deduced from the maximum of $d\rho /dT$ (not shown here) are compatible with the values obtained from the Arrott plots (see Fig. \ref{fig:summary}b below).

Further information can be extracted from the the Gd$_x$Y$_{1-x}$Fe$_2$Zn$_{20}$ $\rho (T)$ data by assuming that the total resistivity of the compound can be written as:
\begin{equation}
\rho (T)=\rho _{0}+\rho _{ph}(T)+\rho _{mag}(T),
\label{eqn:4}
\end{equation}
where $\rho _0$ is a temperature independent, impurity scattering term, $\rho _{ph}$  is the scattering from phonons and $\rho _{mag}$ is scattering associated with the interaction between conduction electrons and magnetic degrees of freedom.
In this series of pseudo-ternary compounds, the high temperature ($T\gg T_{\mathrm{C}}$) phonon contribution, $\rho _{ph}$, should be essentially invariant (due to the very dilute nature of the R ions).
The magnetic contribution to the resistivity, $\rho _{mag}$, will be the combination of contributions from conduction electron scattered by (i) the 4f local moments and (ii) the spin fluctuations of 3d electrons (from Fe sites), both of which should saturate in the high temperature limit.
Based on the analysis above, the high temperature resistivity of the whole series should be similar (modulo an offset) and manifest similar slopes due to the electron-phonon scattering.
This is indeed the case: the data show linearity of $\rho (T)$ above 250 K with the slopes differing by less than 8\%; less than the estimated dimension error (10\%) of these bar-like-shape samples.

The magnetic and disorder contribution to the resistivity can be estimated by (i) removing the geometric error by normalizing the high temperature slope of all $\rho (T)$ plots to that of YFe$_2$Zn$_{20}$ and then (ii) subtracting the $\rho_{Y} (T)$ data from the $\rho$ normalized data.

\begin{figure}
  \begin{center}
  \includegraphics[clip, width=0.45\textwidth]{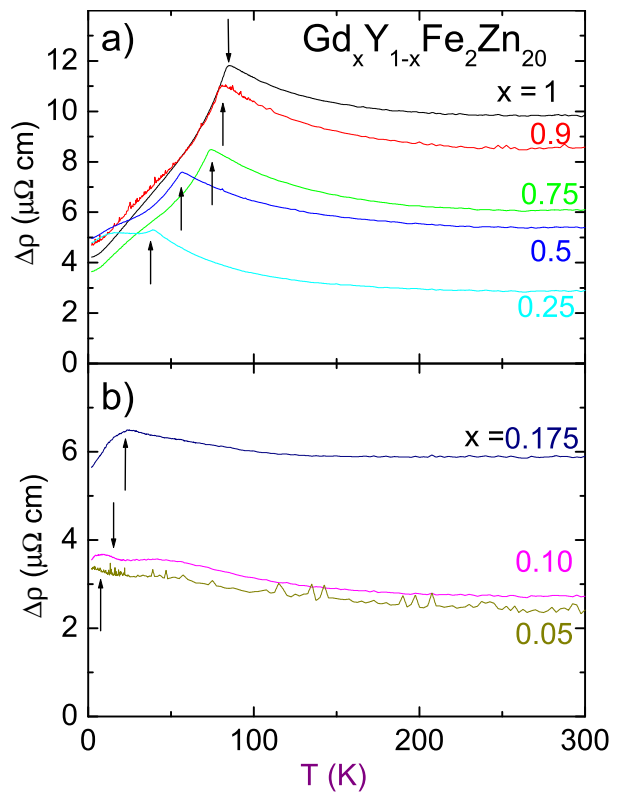}\\
  \caption{Temperature variation of $\Delta \rho $ (see text). The arrows represent $T_{\mathrm{C}}$ determined from Arrott plot analysis of magnetization measurements.}
  \label{fig:DR}
  \end{center}
\end{figure}

The normalized $\rho$ is given as:
\begin{equation}
\rho _{Gd_{x}normalized}=\rho _{Gd_{x}}\cdot \frac{\frac{d\rho _{Gd_{x}}}{dT}\mid _{275K}}{\frac{d\rho _{Y}}{dT}\mid _{275K}}
\label{eqn:5}
\end{equation}
and
\begin{equation}
\Delta \rho =\rho _{Gd_{x}\\normalized}-\rho _{Y}\\.
\label{eqn:6}
\end{equation}
The resulting $\Delta \rho$  will not only show the conduction electron scattering from the 4f local moments, but will also include scattering associated with the interaction between the 4f local moment and 3d electrons, especially near $T_{\mathrm{C}}$.
The temperature dependent $\Delta \rho$ curves for the Gd$_x$Y$_{1-x}$Fe$_2$Zn$_{20}$ compounds are presented in Fig. \ref{fig:DR}.
A pronounced upward cusp is centered about $T_{\mathrm{C}}$  for $x\geq 0.25$.
For $x<0.25$ the loss of spin disorder feature becomes harder (or even impossible) to resolve, but the enhanced scattering above $T_{\mathrm{C}}$ persists.
The decrease of $\Delta \rho$ with T below $T_{\mathrm{C}}$ is a common in ferromagnetic systems and can be explained as the result of a loss of spin disorder scattering of conduction electrons.
On the other hand, the behavior of $\Delta \rho$ above $T_{\mathrm{C}}$ must come from a different conduction electron scattering process. 
A similar feature in $\Delta \rho$ is found in RFe$_2$Zn$_{20}$ (R = Tb-Er) for $T>T_{\mathrm{C}}$\cite{jia_allfe}, but not in isostructural GdCo$_2$Zn$_{20}$ which orders antiferromagnetically at a much lower temperature\cite{jia_nearly_2007}.

\begin{figure}
  \begin{center}
  \includegraphics[clip, width=0.45\textwidth]{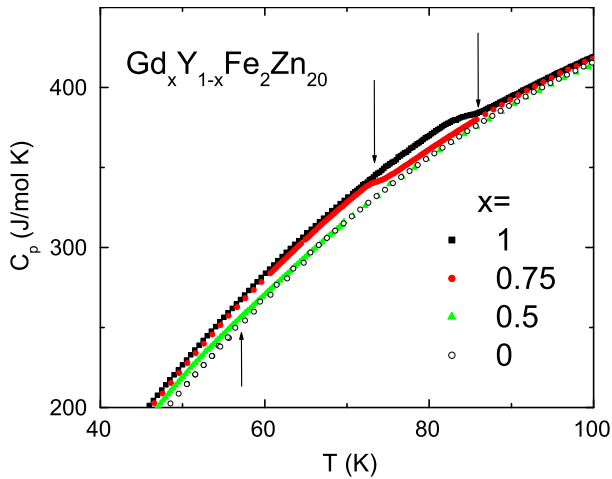}\\
  \caption{Temperature variation of specific heat $C_p$ of the Gd$_x$Y$_{1-x}$Fe$_2$Zn$_{20}$ series for $x$ = 1, 0.75, 0.5 and 0. The arrows represent $T_{\mathrm{C}}$ determined from Arrott plot analyses.}
  \label{fig:Cp}
  \end{center}
\end{figure}

The specific heat of the Gd$_x$Y$_{1-x}$Fe$_2$Zn$_{20}$ compounds (Fig. \ref{fig:Cp}) can be thought of as the sum of the contributions from electronic, vibrational and magnetic degrees of freedom.
To remove the vibrational and electronic parts (at least approximately), the specific heat of YFe$_2$Zn$_{20}$ and LuFe$_2$Zn$_{20}$ were used to estimate the background.
The assumption that YFe$_2$Zn$_{20}$ and LuFe$_2$Zn$_{20}$ closely approx the non-magnetic $C_p$ of the Gd$_x$Y$_{1-x}$Fe$_2$Zn$_{20}$ series is supported by the fact that the difference between the measured $C_p$ of YFe$_2$Zn$_{20}$, LuFe$_2$Zn$_{20}$ and Gd$_x$Y$_{1-x}$Fe$_2$Zn$_{20}$ in the temperature region 20 K higher than $T_{\mathrm{C}}$ is on the order of a percent.
Since LuFe$_2$Zn$_{20}$ has a molar mass closer to that of GdFe$_2$Zn$_{20}$ than YFe$_2$Zn$_{20}$, the combination of $(x)C_{LuFe_2Zn_{20}}+(1-x)C_{YFe_2Zn_{20}}$ is thought to be even closer to the non-magnetic background of $C_{Gd_xY_{1-x}Fe_2Zn_{20}}$.  

\begin{figure}
  \begin{center}
  \includegraphics[clip, width=0.45\textwidth]{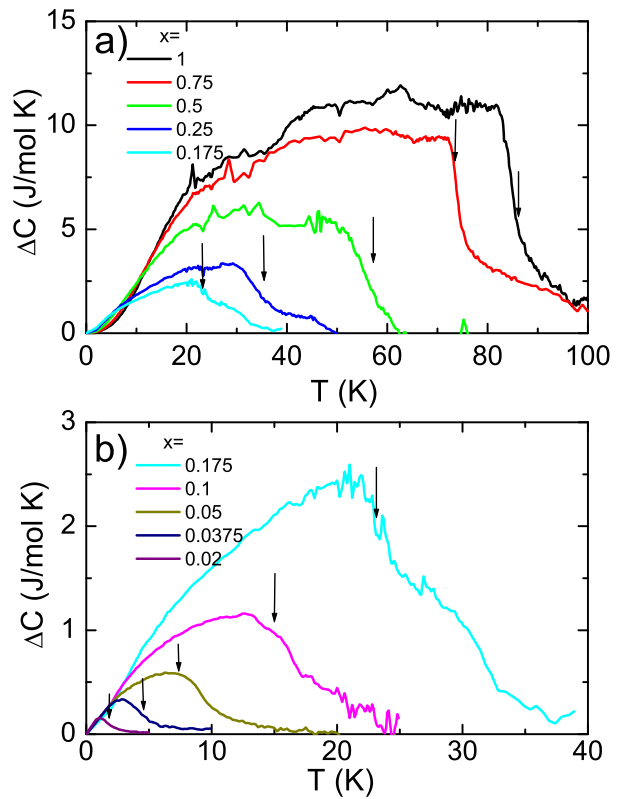}\\
  \caption{Temperature variation of $\Delta C$. The arrows represent $T_{\mathrm{C}}$ values determined from the Arrott analysis of magnetization measurements.}
  \label{fig:DCp}
  \end{center}
\end{figure}

\begin{figure}
  \begin{center}
  \includegraphics[clip, width=0.45\textwidth]{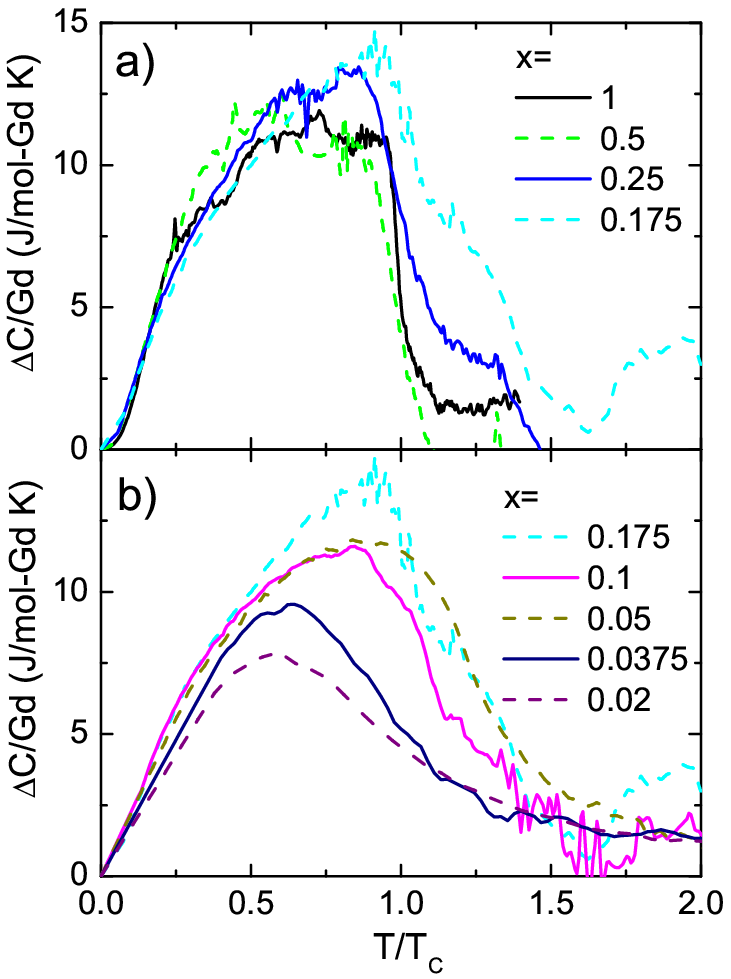}\\
  \caption{$\Delta C/x$ versus $T/T_{\mathrm{C}}$ for representative $x$ values.}
  \label{fig:nDCp}
  \end{center}
\end{figure}

Figure \ref{fig:DCp} shows 
\begin{eqnarray}
\Delta C& = &C_{Gd_xY_{1-x}Fe_2Zn_{20}}\\\nonumber
        & - &(x)C_{LuFe_2Zn_{20}}-(1-x)C_{YFe_2Zn_{20}}
\label{eqn:7}
\end{eqnarray}
for $x\geq 0.175$ (a) and $\leq 0.175$ (b), where the arrows indicate the $T_{\mathrm{C}}$ values determined from the Arrott plot analyses.
The magnetic ordering manifests itself as a broad feature in $\Delta C$ with $T_{\mathrm{C}}$ occurring at, or near, the maximum slope.
Figure \ref{fig:nDCp} shows that this feature persists, relatively unchanged in shape, down to $x=0.1$.
For values of $x<0.1$ the feature broadens further, but is still distinct.
This shape of $\Delta C$ is not unusual for Gd-base intermetallics with ferromagnetic order; for example, a similar feature is seen in GdPtIn ($T_{\mathrm{C}}\sim 68$~K )\cite{morosan_magnetic_2005}.
It should be noted that this $\Delta C$ feature is distinct from that associated with a spin-glass freezing: the maxima all occur at or below $T_{\mathrm{C}}$, whereas a spin glass manifests a broad peak above the freezing temperature\cite{binder_spin_1986}.

The $x$ dependence of the paramagnetic Curie temperature ($\theta _C$), ferromagnetic ordering temperature ($T_{\mathrm{C}}$) and saturated moments per Gd ($\mu _{Sat}$) for each $x$ are shown in Fig. \ref{fig:summary}a, b and c respectively.
The values of the magnetic entropy, estimated by $S_M=\int \frac{\Delta C}{T}dT$ , are shown in Fig. \ref{fig:summary}d.
Both $\theta _C$ and $T_{\mathrm{C}}$  decrease monotonically with $x$.
At first glance, the negative values of $\theta _C$  for $x<0.25$ are unexpected and seem to be in contradiction with the existence of ferromagnetic ground state.
However, the data analyses in Eq. \ref{eqn:1} ignores, the increasingly strong, polarizable background associated with the near Stoner limit conduction electrons at intermediate temperatures.
Furthermore, as shown in Fig. \ref{fig:HM}b, this low temperature effect becomes even more pronounced for small $x$.
Although, as discussed earlier, the uncertainty of $x$ makes the $x$-variation of $\mu _{Sat}$  ambiguous for small $x$, even the large $x$ members of the Gd$_x$Y$_{1-x}$Fe$_2$Zn$_{20}$ series manifest reduced saturated moments.
This is attributed to the induced moment on the 3d electrons, which is anti-parallel to the Gd moment.\cite{jia_nearly_2007}
The magnetic entropy shown in Fig. \ref{fig:summary}d associated with the ordered state is equal to, or slightly larger than, the magnetic entropy associated with the Hund's ground state of Gd$^{3+}$($\textbf {S} = 7/2$).
This fact indicates that the main part of the magnetic specific heat of the series of Gd$_x$Y$_{1-x}$Fe$_2$Zn$_{20}$ is the contribution from the magnetic degrees of freedom of the Gd$^{3+}$ local moments.
The contribution to the magnetic specific heat from the itinerant electrons probably exists, but is, at most, comparable with the measurement uncertainty.

\begin{figure}
  \begin{center}
  \includegraphics[clip, width=0.45\textwidth]{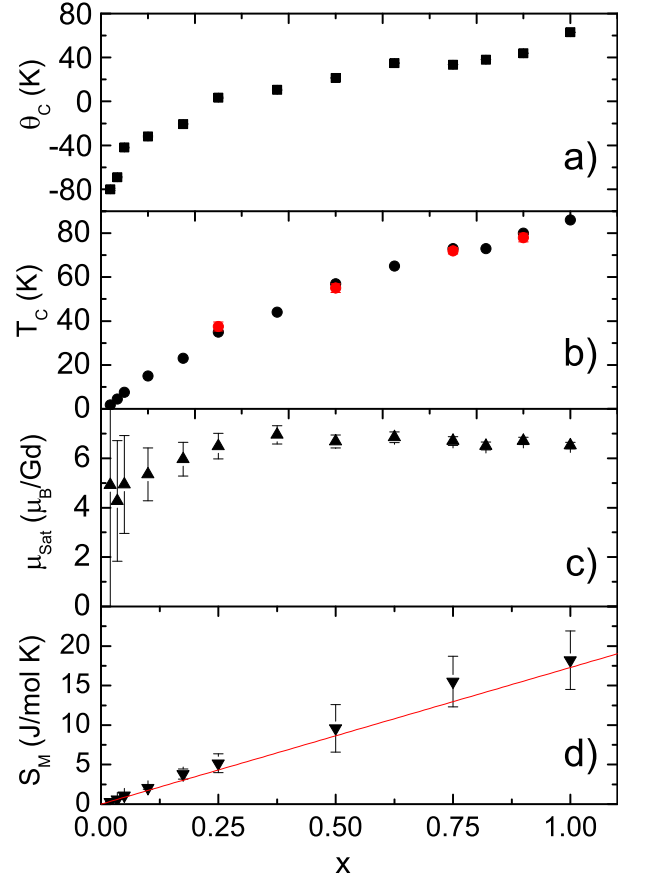}\\
  \caption{(a) Paramagnetic Curie temperature, $\theta _C$, (b) Ferromagnetic ordering temperature, $T_{\mathrm{C}}$, (c) saturated moment per Gd, $\mu _{Sat}$ and (d) magnetic entropy, $S_M$ with respect to $x$ for the Gd$_x$Y$_{1-x}$Fe$_2$Zn$_{20}$ series. The values of $T_{\mathrm{C}}$ in (b) were determined by Arrott plot analyses (black circle) and the resistivity measurements(red circle). The solid line in (d) represents $S_M=xRln8$ ($R$ is gas constant), the magnetic entropy of Gd$^{3+}$ Hund's ground state. The error bars are estimated as 1 \% of the total entropy, $S=\int ^{T_{\mathrm{C}}}_{0}\frac{C_{p}}{T}$.}
  \label{fig:summary}
  \end{center}
\end{figure}

\section{Analysis and Discussion}

For rare earth bearing intermetallics, the interaction between 4f local moments is primarily mediated by means of polarization of the conduction electrons.
Regardless of the details of the mechanism involved in this interaction\cite{de_gennes_RKKY_1962,campbell_indirect_1972}, we emphasis that the 3d electrons from Fe sites act as important mediators of the Gd-Gd interaction in Gd$_x$Y$_{1-x}$Fe$_2$Zn$_{20}$ system. 
In YFe$_2$Zn$_{20}$, the interaction between 3d electrons is not sufficient to split the conduction band but is large enough to make the compound exhibit strongly enhanced paramagnetism. 
When Y$^{3+}$ ions are fully replaced by Gd$^{3+}$ ions, these 3d electrons are polarized by the Gd$^{3+}$ local moments.
 The interaction between 3d electrons assists in stabilizing the splitting of the conduction electron band and enhances the magnetic interaction between Gd$^{3+}$ local moments, resulting in the remarkably high, ferromagnetic transition temperature for GdFe$_2$Zn$_{20}$.
 This physical picture is consistent with the results of the band structure calculation which predicts the Fe induced moment as $0.67 \mu _{B}$ in the ground state of GdFe$_2$Zn$_{20}$\cite{jia_nearly_2007,jia_allfe}.

In order to perform further analysis on the magnetic properties of Gd$_x$Y$_{1-x}$Fe$_2$Zn$_{20}$, a comparision with the binary RCo$_2$ (R = rare earth) intermetallics is useful.
YCo$_2$ and LuCo$_2$ show nearly ferromagnetic behavior while the series of compounds, (Gd-Tm)Co$_2$, with 4f local moments manifest a ferromagnetic ground state\cite{duc_formation_1999,duc_itinerant_1999}.
 In addition to these magnetic similarities, the resemblance between the crystal structure of RT$_2$Zn$_{20}$ and the so-called C-15 Laves structure of RCo$_2$\cite{gschneidner_jr_binary_2006}is noticeable: both rare earth and transition metal ions occupy same unique, single crystallographic sites in same space group: $Fd\bar {3}m$.
The unit cell of the RT$_2$Zn$_{20}$ compounds can be thought of as an expansion of the C-15 Laves phase unit cell via the addition of a large number of Zn ions.

Well-studied for several decades, the series of (Gd-Tm)Co$_2$ has been treated as an example of 4f local moments embedded in a nearly ferromagnetic host: YCo$_2$ or LuCo$_2$.
The so-called s-d model has been employed by Bloch and Lemaire\cite{bloch_metallic_1970} and Bloch et.\ al.\cite{bloch_first_1975} to explain their magnetic properties.
This model was first introduced by Takahashi and Shimizu\cite{takahashi_magnetic_1965} to understand the magnetic properties of alloys of the nearly ferromagnetic transition metal, Pd, with dilute Fe or Co local moment impurities.
In this model, the polarization effect of the local moments on the itinerant electrons is considered in terms of a molecular field.
Motivated by the similarity of the magnetic properties and the crystal structure of RFe$_2$Zn$_{20}$ and RCo$_2$, we applied the s-d model to the Gd$_x$Y$_{1-x}$Fe$_2$Zn$_{20}$ series.

This model considers one magnetic system consisting of two types of spins: one local moment, and the other giving rise to an exchange-enhanced paramagnetic susceptibility.\cite{bloch_metallic_1970} 
 Assuming the interaction between Gd local moments is only via the conduction electrons, we apply this model to Gd$_x$Y$_{1-x}$Fe$_2$Zn$_{20}$. Under an applied field $H$, for $T > T_{\mathrm{C}}$, the magnetization of the Gd local moments and the conduction electrons are:
\begin{equation}
M_{Gd}=(xC_{Gd}/T)(H+n_{Gd-e}M_e)\
\label{eqn:8}
\end{equation}
\begin{equation}
M_e=\chi _{e,0}(H+n_{e-e}M_e+n_{Gd-e}M_{Gd})
\label{eqn:9}
\end{equation}
where $C_{Gd}$ is the Curie constant of the Gd$^{3+}$ local moments; $n_{Gd-e}$, $n_{e-e}$ are molecular-field coefficient representing the interaction between itinerant electrons and Gd$^{3+}$ local moments, and itinerant electrons with themselves, respectively; $\chi _{e,0}$ is the paramagnetic susceptibility without exchange enhancement. The total magnetization of Gd$_x$Y$_{1-x}$Fe$_2$Zn$_{20}$ is the sum of $M_{Gd}$ and $M_e$. It should be noted that when $x = 0$, the total susceptibility reduces to the exchange-enhanced susceptibility:
\begin{equation}
\chi _{e}=\chi _{YFe_2Zn_{20}}=\frac{M_{e}}{H}=\frac{\chi _{e,0}}{1-n_{e-e}\chi _{e,0}}
\label{eqn:10}
\end{equation}
which is simply the Stoner enhanced susceptibility of YFe$_2$Zn$_{20}$.

Assuming that the electronic structure of the conduction band and the position of the Fermi level in the paramagnetic state are the same across the whole Gd$_x$Y$_{1-x}$Fe$_2$Zn$_{20}$ series, from Eqs. \ref{eqn:8}--\ref{eqn:10}, one gets the total susceptibility of Gd$_x$Y$_{1-x}$Fe$_2$Zn$_{20}$
\begin{eqnarray}
\chi _{Gd_xY_{1-x}Fe_2Zn_{20}}&=&\frac {xC_{Gd}}{T-\chi _{YFe_2Zn_{20}}n_{Gd-e}^{2}xC_{Gd}}\\\nonumber
&+&\frac {\chi _{YFe_2Zn_{20}}(T+2n_{Gd-e}xC_{Gd})}{T-\chi _{YFe_2Zn_{20}}n_{Gd-e}^{2}xC_{Gd}}.
\label{eqn:11}
\end{eqnarray}
If one assumes the coupling between the pure spin moment ($\textbf {S} = 7/2$) 
of the Gd$^{3+}$ and the conduction electron spin $\sigma $  ($\sigma =1/2$) to be a Heisenberg exchange interaction, $2J_{0}\vec {S}\cdot \vec {\sigma }$,
 where $J_0$ is the exchange parameter, then the molecular field coefficient 
\begin{equation}
n_{Gd-e}=-J_{0}/(2\mu _{B}^{2}N)
\label{eqn:12}
\end{equation}
where $N$ is the number of the rare earth ions per volume. 

The Gd$_x$Y$_{1-x}$Fe$_2$Zn$_{20}$ system will become ferromagnetic when $\chi _{Gd_xY_{1-x}Fe_2Zn_{20}}$ diverges. Thus, 
\begin{eqnarray}{\label{eqn13}}
T_{\mathrm{C}}& = & \chi _{YFe_2Zn_{20}}(T_{\mathrm{C}})n_{Gd-e}^{2}xC_{Gd}\\\nonumber 
			  & = & x\cdot \chi _{YFe_2Zn_{20}}(T_{\mathrm{C}})\frac{J_{0}^{2}S(S+1)}{3k_{B}N\mu _{B}^{2}}
\end{eqnarray}
where $k_{B}$ is the Boltzmann constant.

\begin{figure}
  \begin{center}
  \includegraphics[clip, width=0.45\textwidth]{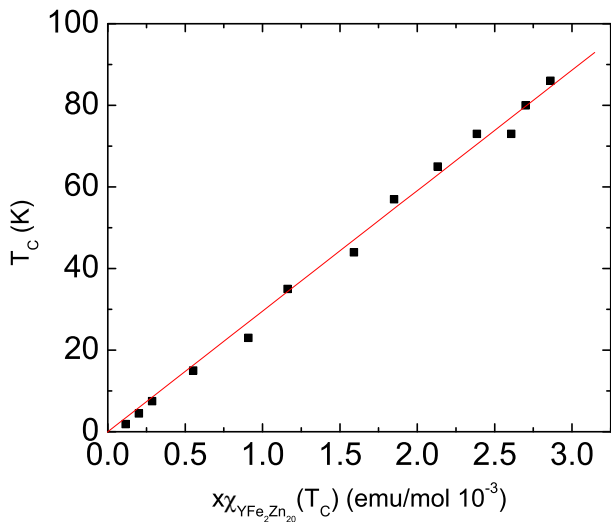}\\
  \caption{$T_{\mathrm{C}}$ of Gd$_x$Y$_{1-x}$Fe$_2$Zn$_{20}$ versus $x\cdot \chi _{YFe_2Zn_{20}}(T_{\mathrm{C}})$. The solid line is linear fit through the origin point}
  \label{fig:sdmodel}
  \end{center}
\end{figure}

Equation \ref{eqn13} reveals that $T_{\mathrm{C}}$ depends on the product of $x$ and $\chi _{YFe_2Zn_{20}}(T_{\mathrm{C}})$, rather than just $x$. This is consistent with Fig. \ref{fig:summary}b showing a nonlinear dependence of $T_{\mathrm{C}}$ on $x$.
Figure \ref{fig:sdmodel} shows that the values of $T_{\mathrm{C}}$ depend linearly on the product $x\chi _{YFe_2Zn_{20}}(T_{\mathrm{C}})$ across the whole series.
From Fig. \ref{fig:sdmodel} the slope equals $2.955\pm 0.0037\times 10^{4}$ K mol/emu and thus $J_0$ can be extracted as $3.96\pm 0.05$ meV.

\begin{figure}
  \begin{center}
  \includegraphics[clip, width=0.45\textwidth]{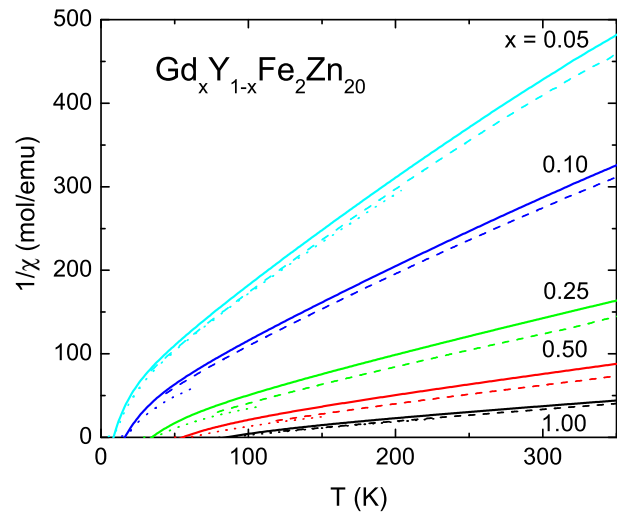}\\
  \caption{$1/\chi $ of Gd$_x$Y$_{1-x}$Fe$_2$Zn$_{20}$ versus $T$ for representative $x$ values. Dot lines: measured under 1 kOe applied filed; dash lines: obtained under high magnetic field; solid lines: calculated results. (See text)}
  \label{fig:fitkia}
  \end{center}
\end{figure}

In addition to the magnetic ordering, this model can also explain the curious temperature dependence of the $1/\chi $ versus $T$ data for the Gd$_x$Y$_{1-x}$Fe$_2$Zn$_{20}$ series. 
Setting $J_{0}=3.96$ meV, one obtains the temperature dependent, total susceptibility of Gd$_x$Y$_{1-x}$Fe$_2$Zn$_{20}$.
The results of $1/\chi _{Gd_{x}Y_{1-x}Fe_{2}Zn_{20}}$ for representative $x$ values are shown as the solid lines in Fig. \ref{fig:fitkia}; whereas the dotted lines and the dash lines present the experimental results under 1 kOe and high magnetic field, representatively.
These calculated results qualitatively reproduce the experimental, temperature dependent susceptibilities, especially their deviation from the Curie-Weiss law close to $T_{\mathrm{C}}$.
It should be noted that the $\chi$ data in Fig. \ref{fig:fitkia} is the full $\chi$ without any subtraction of "non-magnetic" background. 
In this sense Fig. \ref{fig:fitkia}, and the s-d model, appear to treat the magnetization data more fully than the simple assumption behind Eqn. \ref{eqn:1}.  

In addition to the thermodynamic properties discussed above, the feature in $\Delta \rho$ above $T_{\mathrm{C}}$ (Fig. \ref{fig:DR}) is also worth discussing further.
The upward-pointing cusp at $T_{\mathrm{C}}$ of $\Delta \rho(T)$ is associated with the sign change of $d\Delta \rho/dT$ , from negative to positive as the temperature decreases.
 This feature is absent from simple models of $\rho(T)$\cite{craig_transport_1967,fisher_resistive_1968}, based on the models assuming a single lattice of magnetic ions and a single band of conduction electrons.
This theoretical model is over-simplified for Gd$_x$Y$_{1-x}$Fe$_2$Zn$_{20}$, a strongly correlated electron system.
Similar unusual upward cusps in $\Delta \rho(T)$ at $T_{\mathrm{C}}$ were found in the electric transport measurements of RCo$_{2}$\cite{gratz_transport_1995}. 
They were explained by invoking an increasing, non-uniform fluctuating f-d exchange interaction, which provides an increase of spin fluctuation of 3d-electron subsystem as the temperature approaches $T_{\mathrm{C}}$ in paramagnetic state, which in turn leads to increased conduction electron scattering.
The Gd$_x$Y$_{1-x}$Fe$_2$Zn$_{20}$ system (and indeed the other RFe$_2$Zn$_{20}$ compounds\cite{jia_allfe}) present another, clear example of this behavior.
\section{Summary}

We presented a set of data including magnetization, electrical transport and specific heat, measured on flux-grown single crystals of Gd$_x$Y$_{1-x}$Fe$_2$Zn$_{20}$.
We found that the series order ferromagnetically above 1.85 K for $x \geq  0.02$.
The variation of $T_{\mathrm{C}}$ with respect to x, as well as the curious temperature dependent magnetic susceptibilities, are well explained by a modifucation of the s-d model based on molecular field approximation.
The enhanced electronic scattering found for $T\geq T_{\mathrm{C}}$ is qualitatively consistent with this model and similar to that found for the related RCo$_2$ compounds.

\begin{acknowledgments}
The authors thank J. Frederich and M. Lampe for growing some of the compounds, F. Laabs for EDS measurements, E. D. Mun, A. Safa-Sefat and J. Schmalian for helpful discussions.
Ames Laboratory is operated for the U.S. Department of Energy by Iowa State University under Contract No. DE-AC02-07CH11358.
This work was supported by the Director for Energy Research, Office of Basic Energy Sciences.

\end{acknowledgments}

\end{document}